\documentclass{article}
\usepackage[utf8]{inputenc}
\usepackage[margin=1in]{geometry}
\usepackage{amsmath}
\usepackage{physics}
\usepackage{graphicx}
\usepackage{amssymb}
\usepackage{color}
\usepackage{authblk}

\newcommand{\SFF}{\text{SFF}}
\newcommand{\bbeta}{{\boldsymbol \beta}}

\begin{document}
\title{Spontaneous Symmetry Breaking, Spectral Statistics, and the Ramp}
\author[1]{Michael Winer}
\author[2]{Brian Swingle}
\affil[1]{Condensed Matter Theory Center and Joint Quantum Institute,
Department of Physics, University of Maryland, College Park, MD 20742, USA}
\affil[2]{Brandeis University, Waltham, Massachusetts, USA 02454}
\maketitle 
\begin{abstract}
Ensembles of quantum chaotic systems are expected to exhibit energy eigenvalues with random-matrix-like level repulsion between pairs of energies separated by less than the inverse Thouless time. Recent research has shown that exact and approximate global symmetries of a system have clear signatures in these spectral statistics, enhancing the spectral form factor or correspondingly weakening level repulsion. This paper extends those results to the case of spontaneous symmetry breaking, and shows that, surprisingly, spontaneously breaking a symmetry further enhances the spectral form factor. For both RMT-inspired toy models and models where the symmetry breaking has a description in terms of fluctuating hydrodynamics, we obtain formulas for this enhancement for arbitrary symmetry breaking patterns, including $Z_n$, $U(1)$, and partially or fully broken non-Abelian symmetries.
\end{abstract}

\section{Introduction}

This paper studies the statistical properties of energy levels of chaotic quantum systems exhibiting spontaneous symmetry breaking (SSB). The phenomenon of SSB can occur whenever a system possesses a symmetry and a suitable thermodynamic limit. SSB is said to occur at a given energy density if, after first taking the thermodynamic limit, a vanishingly small symmetry breaking perturbation in the Hamiltonian leads to a non-symmetric equilibrium state (reviews include \cite{Beekman_2019,Hidaka_2020}). Because the system's symmetry constrains the structure of the energy spectrum, SSB also manifests as a certain reorganization of the energy levels as a function of energy density. It is the purpose of this paper to understand this reorganization and how it crosses over into the symmetry unbroken case after the Thouless time in a finite size system.

Typically, symmetries are unbroken at high energy density and may be broken as the energy density approaches the edges of the spectrum. We focus on quantum chaotic systems which have symmetric random-matrix-like energy levels at high energy density~\cite{bohigas1984chaotic,berry1977level,doi:10.1063/1.1703775,haake2010quantum}. As the energy density is lowered, the occurence of SSB can then be understood as a breaking of ergodicity in the thermodynamic limit. Our theory quantitatively explains how ergodicity and symmetry are restored at finite system size from the point of the view of the energy spectrum. There have also been a few other studies of the interplay of quantum chaos, eigenstate thermalization, and spontaneous symmetry breaking including~\cite{Zhao_2014,Fratus_2015,fratus2017eigenstate}; see~\cite{D_Alessio_2016} for a review of notions of quantum chaos.

Because symmetry restoration is a long-time process, the theory must deal with special slow dynamics associated with the order parameter of the broken symmetry for which a hydrodynamic-like effective theory is the right description~\cite{crossley2017effective,Glorioso_2017,Grozdanov_2015,Kovtun_2012,Dubovsky_2012,Endlich_2013}. In previous work~\cite{winer2020hydrodynamic}, we showed how the quantum field theory formulation of fluctuating hydrodynamics could be adapted to predict a random-matrix-like spectral form factor at late times and to compute corrections to finite time corrections to random matrix theory (RMT) due to slow modes. The present paper can be viewed as an extension of the earlier theory to systems with the additional physics of spontaneous symmetry breaking\cite{Lallouet_2003}\cite{Hurtado_2011}.

We now describe the setup in more detail. A system has symmetry group $G$ if (1) $G$ acts on the Hilbert space of the system by some faithful (but typically reducible) representation $\mathcal{U}$ and (2) the Hamiltonian of the system commutes with every representative, $\mathcal{U}(g) H = H \mathcal{U}(g)$. We focus on systems where the representation is unitary and linear. Given a maximal commuting set of elements of $G$, we can find simultaneous eigenstates of $H$ and this maximal commuting set. The Hamiltonian breaks up into blocks labelled by the irreducible representations (irreps) of $G$. 

In the simplest quantum chaotic case, each irrep block will consist of a number of copies (equal to the dimension of the irrep) of a random matrix~\cite{bohigas1984chaotic,berry1977level,doi:10.1063/1.1703775,haake2010quantum}, and the matrices for different irreps will be independent. We quantify the random character of these matrices using a filtered form of the spectral form factor (SFF)~\cite{brezin1997spectral} which zooms in on a particular energy density. Filtering is important because SSB is an energy-density-dependent phenomenon. The spectral form factor with filter function $f$ is
\begin{equation}
    \SFF(T,f) = \overline{ | \text{Tr}[U(T) f(H)]|^2},
\end{equation}
where $U(T) = e^{-i H T}$ is the time evolution operator and overline denotes a disorder average over an ensemble of Hamiltonians where each representative is symmetric. We will say more about this averaging shortly; it is necessary to render the SFF a smooth function of time $T$.
\begin{figure}
\centering
\includegraphics[scale=0.5]{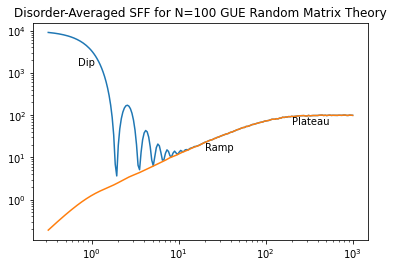}
\caption{SFF (blue) of a random matrix exhibiting the dip-ramp-plateau structure. The connected SFF only exhibits the ramp and plateau.}
\end{figure}
It is also useful to define the connected and disconnected contributions to the SFF. The disconnected part, 
\begin{equation}\SFF_{\text{dis}}(T,f)=|\overline{\text{Tr}[U(T) f(H)]}|^2,
\end{equation}
is distinguished by the full SFF because the squaring happens after the averaging. The connected part is the difference
\begin{equation}\SFF_{\text{con}}(T,f)=\SFF(T,f)-\SFF_{\text{dis}}(T,f)
\end{equation}
It can be shown that both the connected and disconnected parts are positive definite. In the standard dip-ramp-plateau picture, the dip comes from the disconnected component, and the ramp and plateau come from the autocorrelation found in the connected component of the SFF.

For a quantum chaotic system with no symmetry, the expectation is that, after a Thouless time the SFF will agree with the random matrix result \cite{bohigas1984chaotic,berry1977level,doi:10.1063/1.1703775,haake2010quantum}. Up to the Heisenberg time, which is proportional to the level density, the RMT result is~\cite{mehta2004random} the linear ramp,
\begin{equation}
    \SFF(T,f) = \int dE f^2(E) \frac{T}{\pi \bbeta},
\end{equation}
where $\mathfrak{b} =1,2,4$ is the Dyson index. If we choose $f$ to be a Gaussian filter of the form $f = \exp\left( -\frac{(E-E_0)^2}{4\sigma^2}\right)$, then the result is
\begin{equation}
    \SFF(T,f) = \frac{\sqrt{2\pi} \sigma T}{\pi \bbeta}
\end{equation}
provided $E_0$ sits within the spectrum of $H$. If instead we have a system with $G$ symmetry which is unbroken at energy $E_0$, then the Gaussian filtered SFF will be
\begin{equation}
    \SFF(T,f) = \sum_{R} d_R^2 \frac{\sqrt{2\pi} \sigma T}{\pi \bbeta},
\end{equation}
where the sum is over irreps appearing in the spectrum and $d_R$ is the dimension of the representation. Each irrep block is composed of $d_R$ identical copies of an independent random matrix (one for each $R$), so the $d_R^2$ factor arises because all the subblocks are perfectly correlated.

The new feature associated with SSB at energy $E_0$ is that the different irrep blocks will no longer be effectively independent. At large but finite system size, the different blocks will be strongly correlated because the splittings between blocks must be suppressed in order to have SSB. This additional correlation will cause the SFF to take a larger value at early to intermediate times. Then at fixed system size (and for typical forms of SSB), the system will crossover to the unbroken behavior at very long time provided the order parameter fluctuates rapidly compared to the Heisenberg time of each block. This paper reports a theory of this physics as manifested in the filtered SFF.

The remainder of the paper is organized as follows. In section \ref{sec:Zn} we will consider the case of spontaneously broken $Z_n$ symmetry in various toy models, obtaining analytic and numerical results in excellent agreement. Next in section \ref{sec:discreteG} we will do the same for more general finite groups $G$. Next we will consider a hydrodynamic calculation of the symmetry-breaking SFF in sections \ref{sec:AbelianHydro} and \ref{sec:NonAbelianHydro}, focusing first on the Abelian and then on general Lie Groups.

\section{$Z_n$ SSB With Zero Spatial Dimensions}
\label{sec:Zn}

We will start with the simplest possible case of discrete SSB: the case of a $Z_n$ symmetry and a charge-$1$ order parameter. Since any discrete abelian group is a product of $Z_n$ factors (with possibly different $n$s), this case captures most of the interesting physics of discrete abelian SSB. We also restrict attention to the case of zero spatial dimensions, to the reader should have in mind a cluster or similar sort of system where a large number of degrees of freedom can interact without geometric restrictions.

The the charge-$1$ order parameter is described by a basis $\ket \phi \in \Phi$ with $\phi$ an integer from $0$ to $n-1$. We take the other degrees of freedom to be described by a state $\psi$ in an $N$-dimensional Hilbert space $\Psi$. A state in the total Hilbert space will be an $nN$-dimensional superposition of states of the form $\ket \phi \otimes \ket \psi$. In this section, we will require that $\ket \psi$ transform trivially under $G$. We briefly consider more general behavior in appendix \ref{app:IntCharge}.

The Hamiltonian is built from a collection of operators $H_k$ acting on $\Psi$ that are associated with transitions of the order parameter from sector $\phi$ to sector $\phi+k$. Using the shift operator $M_k$ defined as $M_k |\phi\rangle = |\phi + k\rangle$, we write this decomposition as
\begin{equation}
\begin{split}
    H=\sum_k M_k\otimes H_k\\
    (M_k)_{ij}=\delta_{i,j+k},
\end{split}
\end{equation}
with the arguments of the delta function in $M_k$ all taken mod $n$. Hermiticity of $H$ requires that $H_k^\dagger = H_{n-k}$ since $M_k^\dagger = M_{n-k}$. For instance, the $n=4$ Hamiltonian written out in block matrix form is
\begin{equation}
\begin{split}
    H=\begin{pmatrix}
    H_0&H_3&H_2&H_1\\
    H_1&H_0&H_3&H_2\\
    H_2&H_1&H_0&H_3\\
    H_3&H_2&H_1&H_0\\
    \end{pmatrix}\\
    \text{with } H_0=H_0^\dagger, H_1=H_3^\dagger, H_2=H_2^\dagger.\\
\end{split}
\end{equation}

At this point, we must ask what sorts of matrices $H_0$ and $H_k$ make good models of the sorts of systems we see in real life. As a simple model, consider the case where each $H_k$ is chosen independently consistent with the constraints imposed by hermiticity. This should be a reasonable description of the spectral properties of generic chaotic systems after all other modes have decayed. Each block has matrix elements with variance $J_k^2/N$, and the physics of SSB is modeled by the condition $J_0\gg J_{k\neq 0}$. This leads to a diffusive motion in the order parameter space.  

The $H_k$ could also have more structure. For example, one could have $H_k=f_k(H_0)$ for some simple slowly-varying matrix polynomial function $f$. A salient case is where $f_0$ is roughly constant over the energy range of $H_0$, and is $0$ unless $k=\pm 1$.  In this world, the Hamiltonian can be roughly written as $I \otimes H_0 + K \otimes f(H_0)$, where $K_{ij}=\delta_{i,j+1}+\delta_{i,j-1}$. We can diagonalize $H_0$ and $K$ to diagonalize this matrix. Given an eigenstate $\ket \psi$ of $H_0$ with $H_0\ket \psi=E_\psi \ket \psi$ and an eigenstate $\ket q$ of $K$ with $K\ket q=E_q \ket q$ the energy of $\ket q \otimes\ket \psi$ is $E_\psi+f(E_\psi)E_q$. Especially when $n$ is large, it makes sense to talk about a dispersion relation depending on $q$. We can think of the state as a particle with internal degrees of freedom propagating ballistically in order parameter space. Many real-life SSB systems with a large or continuous broken symmetry exhibit such ballistic propagation combined with effects of the purely random model.

\subsection{The SFF With a Purely Random Kinetic Term}

Consider first the case of purely random $H_k$. To calculate the SFF, it is convenient to work with two copies of the system with total Hamiltonian
\begin{equation}
H_{\text{tot}}=H_{\text{sys}}\otimes I-I\otimes H^*_{\text{sys}}.
\end{equation}
The SFF of a single copy of the system is then
\begin{equation}
    \text{SFF}(T,1)=\overline{\tr \exp(-iH_{\text{tot}}T)}.
\end{equation}
Note that the complex conjugation in the definition of $H_{\text{tot}}$ is added for interpretational convenience; it means that $H_{\text{tot}}$ has the nice property of annihilating a maximally entangled state made of Bell pairs.
\begin{figure}
\centering
\begin{tabular}{cc}
\includegraphics[scale=0.5]{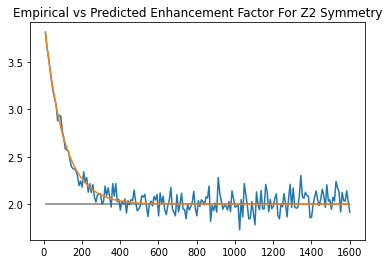}&
\includegraphics[scale=0.5]{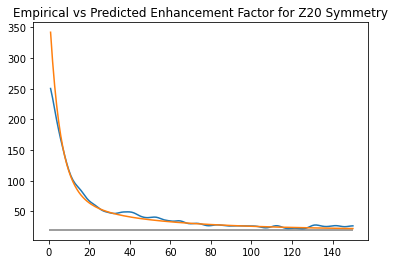}\\
\end{tabular}
\caption{The predicted (orange) vs displayed (blue) enhancements for two $Z_n$-symmetric Hamiltonians. The pure RMT prediction for a matrix with that symmetry group is in gray.}
\label{fig:ZnGraph}
\end{figure}

As a reminder, if all the $H_{k \neq 0}$ are set to zero, then the order parameter is frozen and the SFF is controlled by the diagonal $H_0$ blocks. $H_0$ is a random matrix, so its SFF is given by the random matrix result, but since each block is identical and there are $n$ blocks, the total SFF of the system is $n^2$ times the random matrix result. Nonzero $H_{k \neq 0}$ blocks cause the order parameter to fluctuate. We want to calculate the modifications to the SFF by summing over quantum trajectories of the order parameter. This calculation is viewed as a kind of path integral calculation on the doubled system with Hamiltonian $H_{\text{tot}}$. The averaging operation discussed in the introduction corresponds here to averaging over the elements of the $H_k$s.

The SFF is essentially an return probability summed over initial states. Let's take a moment to think about what sort of formula we should expect for this sum. As with other ramp-related quantities, it is helpful to juxtapose it with Schwinger-Keldysh/CTP contour $e^{-\beta H}e^{iTH}e^{-iTH}$. We can generally factor the Hilbert space into slow degrees of freedom $\Phi$ and fast degrees $\Psi$. In this paper $\Phi$ is the order parameter of a symmetry and the various $\Phi$ sectors will be related by that symmetry. But $\Phi$ could just as well include conserved quantities, gauge bosons, or fermion degrees of freedom with a chiral symmetry.

At any point along the contour we have some density matrix $\rho_{\Phi\Psi}\propto e^{-\beta H}$ indicating the density for both the Goldstone mode $\phi$ and the microscopic degrees of freedom $\psi$ (equivalently, this can be viewed as a state of the two-replica system). If we trace out the $\Psi$ modes, $\rho_{\Phi}$ evolves by multiplication by a time-dependent superoperator, $e^{-\text{Trans}(E) T}$, which is generated from a sort of transfer matrix $\text{Trans}(E)$ that acts on two replicas of the order parameter space. More precisely,
\begin{equation}
    \rho_{\Phi}(T)=\int dE e^{-\text{Trans}(E)T} P_E\rho_{\Phi}(T)
\end{equation}
where $P_E$ projects down to an energy window around $E$.

If it weren't for the symmetry relating the different $\phi$s, the superoperator $\text{Trans}(E)$ would approximately annihilate doubled-system states $\ket{\phi_1}\ket{\phi_2}$ with $\phi_1\neq \phi_2$, and we could essentially replace
\begin{equation}
    \text{Trans}(E)_{\phi_1 \phi_2 \phi'_1 \phi'_2}\rightarrow \delta_{\phi_1 \phi'_1} \delta_{\phi_2 \phi'_2} \text{Trans}(E)_{\phi_1 \phi'_1}.
\end{equation} 
This is essentially the case considered in \cite{winer2020hydrodynamic}. But here the symmetry allows for constructive interference, and we need to trace over $\text{Trans}(E)$ in its full glory.

If the matrix elements of $H_0$ and $H_{k \neq 0}$ are independent random numbers, then there are only two perturbative processes that contribute to the path integral and thus to $\text{Trans}(E)$. The first is when both systems go from $\psi_1$ to $\psi_2$ using the same $H_k$. The second is when one of the two systems goes from $\psi_1$ to $\psi_2$ to $\psi_1$, using $H_k$ and $H_k^\dagger$. The both-sides-jump perturbation is parameterized by the two times $t_1,t_2$ of the two jumps, and by the energies $E_{\text{final}}$ after the jump. Disorder-averaging over the matrix elements and converting the sum over final states into an integral, the total amplitude for this process to happen once after time $T$ is
\begin{equation}
\sum_{E_{\text{final}}}\int dt_1dt_2 |H_k\textrm{ rms matrix element}|^2e^{i(E_{\text{final}}-E_{\text{init}})(t_2-t_1)}= r_k(E_{\text{init}}) T
\end{equation}
where the rate $r_k$ is
\begin{equation}
    r_k(E) = 2\pi |H_k\textrm{ rms matrix element}|^2 \rho(E)
\end{equation}
with $\rho(E)$ the density of states; this is just the Fermi's golden rule. For the $\psi_1\to\psi_2\to\psi_1$ processes, we have the condition $t_2>t_1$, which produces a factor of $\frac 12$ relative to when the two jumps happen on separate contours.

Including the associated order parameter dynamics, we get a transfer matrix of the form
\begin{equation}
\text{Trans}(E)=  \sum_{k\neq 0} r_k(E)\left(  I \otimes I - M_k\otimes M_k \right) =  \frac{1}{2} \sum_{k\neq 0} r_k(E)\left( 2 I \otimes I - M_k\otimes M_k - M_{-k} M_{-k} \right)
\label{eq:transferEq}
\end{equation}
The second equality follows from $r_{k} = r_{n-k}$ and $M_{-k} = M_{n-k} = M_k^\dagger$. Note that this is a tensor product of different spaces, two factors of $\Phi$, rather than a factor of $\Phi$ and a factor of $\Psi$. This means that the transfer matrix has four $\Phi$ indices, i.e. it is a superoperator. The enhancement factor to the SFF is just $\int dE f^2(E) \tr e^{-\text{Trans}(E) T}$. We can see this prediction borne out in Figure~\ref{fig:ZnGraph}.

Note that the late time enhancement of the SFF follows from the number of zero modes of $\text{Trans}(E)$. The general spectrum of the transfer matrix is obtained from states of the form
\begin{equation}
    \sum_{\phi \phi'} e^{i q \phi + i q' \phi'} \ket \phi \ket{\phi'}, 
\end{equation}
where $q,q'$ are $\frac{2\pi}{n}$ times an integer $\in \{0, \cdots, n-1\}$. The $q,q'$ state has eigenvalue
\begin{equation}
    \sum_{k\neq 0} r_k (1 - \cos k (q+q') ).
\end{equation}
The set of zero modes is given by $q=-q'$, hence, there are $n$ zero modes of $\text{Trans}(E)$. This implies that the late time SFF enhancement is $n$, as expected in a situation where the symmetry has been restored.

\subsection{The SFF With A Mixed Kinetic Term}

\begin{figure}
\centering
\includegraphics[scale=0.5]{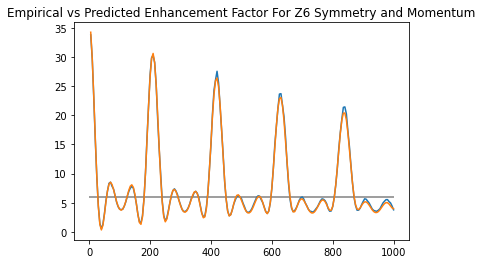}
\caption{Each $H_0$ is a 500 by 500 matrix, with consecutive blocks connected by $c_1R+c_2I$,  where $R$ is a random complex matrix,  $c_1$=0.02, and $c_2$=0.03. The numerical results are in blue, the analytic predictions are red. The pure RMT prediction for a matrix with that symmetry group is in gray.}
\label{fig:ZnMomentum}
\end{figure}

The effect of adding in a slowly varying kinetic term like $H_k = f_k(H_0)$ is to add in additional processes which can contribute to transfer matrix. More precisely, let's break our Hamiltonian into
\begin{equation}
H_{\text{sys}}=\sum_k M_k \otimes H_k  + \sum_k M_k \otimes f_k(H_0),
\end{equation}
with the $H_k$s fully independent of $H_0$ and the $f_j$s reasonably slowly varying analytic functions of small enough value to be treated by perturbation theory. We will get an additional contribution to the transfer matrix of
\begin{equation}
\text{Trans}(E)_{\text{kinetic}} = - i \sum_k f_k(E) \left( M_k\otimes I-I \otimes M_k \right),
\label{eq:transfer2}
\end{equation}
which is just the amplitude of a single jump caused by the perturbation $f_j$. Here again the tensor product refers to two factors of $\Phi$. We see this prediction confirmed in Figure~\ref{fig:ZnMomentum}.

\section{General Discrete SSB}
\label{sec:discreteG}

In this section we will generalize from the simple Abelian case from last section and consider an arbitrary (finite) non-Abelian symmetry group $G$. Apart from its intrinsic interest, one motivation for considering $Z_n$ SSB is that we can approach $U(1)$ SSB by taking the limit $n\rightarrow \infty$. Unfortunately, there is no way to make arbitrarily good approximations of continuous non-Abelian groups with finite groups. Nonetheless, such discrete non-Abelian groups do appear in nature and can be spontaneously broken. A standard example is a valence bond solid, which preserves the $SU(2)$ spin symmetry but breaks the crystal point group symmetry.

We will start with some discrete symmetry group $G$, and some order parameter $\phi\in \Phi$. It will be helpful to think of $\phi$ not necessarily as a number, but as a thing which transforms under $G$. For instance in a ferromagnet, $\phi$ is a unit vector on the unit sphere $S^2$ indicating the direction of polarization. It transforms under the rotation group $G=SO(3)$. In effect, each element $g\in G$ induces a function $f_g(S^2)\to S^2$, where $f_{g_1}\circ f_{g_2}=f_{g_2g_1}$. Mathematicians call this a group action~\cite{gAction}. We say that $\Phi$ is the orbit of $\phi$ under $G$. There is some (possibly trivial) subgroup $G'$ which maps $\phi$ to itself. This is called the stabilizer of $\phi$, and we have $|G'||\Phi|=|G|$. 

In addition to the order parameter space $\Phi$,  we again we will also have an 'internal' state $\ket \psi \in \Psi$, which transforms trivially under $G$. Once again we will set $\Phi$ to be $N$-dimensional. There is again some Hamiltonian random $H_0$ (variance $J_0^2/N$) within each of those $\phi$ subspaces, and a number of different matrices $\{H_i\}$ connecting different sectors. Let's again choose each element independently, with variance $J_i^2/N$, and where $J_{i}\ll J_0$. Because of the symmetry, each of the $H_i$s will show up in more than one place. 

Consider the example of the group $D_4$, which can be realized as the symmetry group of a square. This is a discrete group with $8$ elements generated by $90^{\circ}$ counterclockwise rotations ($a$) and reflections about one diagonal ($b$). These obey the relations $a^4=b^2 =e $ ($e$ is the identity) and $ab = b a^{-1}$. The center is $\{e,a^2\}$. The group has four one-dimensional irreps and one two-dimensional irrep. 

The order parameter corresponds to a choice of one corner of the square, so it takes $4$ values. Suppose that the only allowed jump of the order parameter corresponds to the rotation $a$ and the inverse rotation $a^{-1} = a^3$. Let the corresponding operators acting on $\Psi$ be $H_a$ and $H_{a^3}$. Invariance under the reflection $b$ requires $H_a = H_{a^3}$. The block matrix form of the system Hamiltonian is thus
\begin{equation}
\begin{split}
H=\begin{pmatrix}
H_0&H_a&0&H_a\\
H_a&H_0&H_a&0\\
0&H_a&H_0&H_a\\
H_a&0&H_a&H_0
\end{pmatrix}
\end{split}
\end{equation}

More generally, the system Hamiltonian is given by
\begin{equation}
H_{\text{sys}}=I \otimes H_0+\sum_i M_i\otimes H_i+M_i^\dagger\otimes H_i^\dagger
\label{eq:HRandom}
\end{equation}
where the $M_i$s are matrices which are mostly $0$s with a few $1$s connecting pairs of $\phi$s related by the same group action.  More precisely, the $M_i$s are indexed by $\Phi^2/G$. That is two ordered pairs of $\phi$s share the same $M_i$ if their pairs are connected by an element of $G$.

In the non-Abelian case, we only consider the situation where the $H_i$ matrices are completely random and uncorrelated with $H_0$ and each other. In this case, the walk over the order parameter space is a random walk. This is a good model for, say, a particle stuck in one of $k$ identical wells (with enough internal structure to be chaotic), which jumps between wells using diffusive Poisson-frequency instantons. If the wells are placed on the vertices of a $k$-gon, the relevant group would be dihedral group $D_k$. If they were the vertices of a $k-1$-simplex, the relevant group would be the symmetric group $S_k$. This paradigm can be mathematically modeled.

\subsection{The SFF With a Purely Random Kinetic Term}

\begin{figure}
\centering
\begin{tabular}{cc}
\includegraphics[scale=0.5]{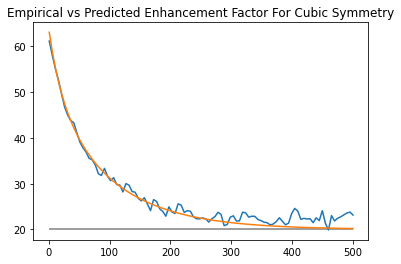}&
\includegraphics[scale=0.5]{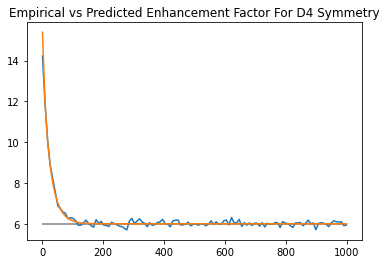}\\
\includegraphics[scale=0.5]{plots/Z2.png}&
\includegraphics[scale=0.5]{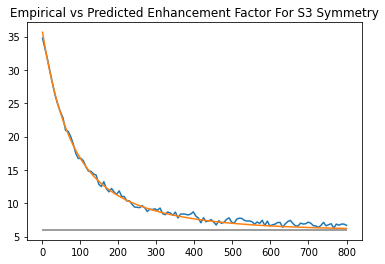}
\end{tabular}
\caption{The predicted (orange) vs displayed (blue) enhancements for a variety of group actions: Cubic Symmetry acting on the vertices of a cube, $D_4$ acting on the corners of a square,  $Z_2$ acting on two points, $S_3$ acting by multiplication on elements of $S_3$.}
\label{fig:nonAbelian}
\end{figure}
As in the Abelian case, to calculate the SFF, we want to have two copies of the system with total Hamiltonian
\begin{equation}
H_{\text{tot}}=H_{\text{sys}}\otimes I-I\otimes H^*_{\text{sys}}
\end{equation}
The SFF of a single copy of the system is still
\begin{equation}
    \text{SFF}(T,1)=\overline{\tr \exp(-iH_{\text{tot}}T)}.
\end{equation}
The analysis proceeds as in the Abelian case, in particular, we still have a path integral picture and treat the order parameter jumps in perturbation theory.


For each jump type $M_i$, there is a corresponding rate $r_i$. The transfer matrix is
\begin{equation}
\begin{split}
\text{Trans}(E)= \frac 12 \sum r_i(E)\left( \{ M_i, M_i^T \} \otimes I + I \otimes \{ M_i, M_i^T\} - 2M_i\otimes M_i-2M_i^T\otimes M_i^T\right)\\
= \frac 12 \sum r_i(E) (M_i\otimes I-I\otimes M_i^T)(M_i^T\otimes I-I\otimes M_i)+(M_i^T\otimes I-I\otimes M_i)(M_i\otimes I-I\otimes M_i^T),
\end{split}
\label{eq:transferEqNonAbelian}
\end{equation}
where we used the fact that the $M_i$ have real matrix elements. The enhancement factor to the SFF is $\tr e^{-\text{Trans}(E) T}$. Figure \ref{fig:nonAbelian} shows the predicted and realized enhancements for various choices of $G$ and $\Phi$. In appendix \ref{app:longTime} we show that at long times this gives the RMT result consistent with symmetry $G$.

\section{Abelian SSB Hydro and the SFF}
\label{sec:AbelianHydro}

We now turn to the description of SFFs in systems with spontaneously broken continuous symmetries. We begin with the Abelian case, focusing again on the simplest case of $U(1)$ symmetry and a charge $1$ order parameter. Here the full power of the corresponding Schwinger-Keldysh hydro effective theory is revealed, so we first review that theory and then describe its modification to treat the SFF as in~\cite{winer2020hydrodynamic}.

\subsection{Quick Review of Hydrodynamics}

At the broadest level, hydrodynamics is the program of creating effective field theories (EFTs) for systems based on the principle that macro-physics should be driven primarily by conservation laws. We will utilize the technology of the CTP formalism. For an accessible introduction, see~\cite{glorioso2018lectures}, and for more details see ~\cite{crossley2017effective,Glorioso_2017}. Additional information about fluctuating hydrodyamics can be found in~\cite{Grozdanov_2015,Kovtun_2012,Dubovsky_2012,Endlich_2013}. We first discussed the application of hydro EFTs for non-SSB spectral statistics in~\cite{winer2020hydrodynamic}.

The CTP formalism starts with the following partition function of a Schwinger-Keldysh contour: 
\begin{equation}
    Z[A^\mu_1(t,x),A^\mu_2(t,x)]=\tr \left( e^{-\beta H} \mathcal{P} e^{i\int dt d^d x A^\mu_1j_{1\mu}} \mathcal{P} e^{-i\int dt d^d x A^\mu_2 j_{2\mu}}\right).
\end{equation}
For $A_1=A_2=0$, this is just the thermal partition function. Differentiating with respect to the $A$s generates insertions of the conserved current density $j_\mu$ along both legs of the Schwinger-Keldysh contour. Thus $Z$ is a generator of all possible contour ordered correlation functions of current operators.

One way to enforce the conservation law $\partial^\mu j_{i\mu}=0$ is to require 
\begin{equation}
    \begin{split}
         Z[A^\mu_1(t,x),A^\mu_2(t,x)]=\int \mathcal D \phi_1\mathcal D \phi_2 \exp\left(i\int dt dx W[B_{1\mu},B_{2\mu}]\right),\\
         B_{i\mu}=\partial_\mu \phi_i+A_{i\mu}.
    \end{split}
    \label{eq:AbelianB}
\end{equation}
Here the fields $\phi_i$ have been ``integrated in'' and represent the slow fluctuating modes of the system. This presentation of $Z$ guarantees $\partial_{\mu}\partial_{A_{i\mu}}Z=0$ as an operator identity. 

The functional $W$ is not arbitrary. The key assumption is that the hydro action $W$ is local. Moreover, when expressed in terms of
\begin{equation}
\begin{split}
    B_a=B_1-B_2,\\
    B_r=\frac{B_1+B_2}{2},
\end{split}
\end{equation}
there are several constraints which follow from unitarity:
\begin{itemize}
    \item $W$ terms all have at least one power of $B_a$, that is $W=0$ when $B_a=0$,
    \item Terms odd (even) in $B_a$ make a real (imaginary) contribution to the action,
    \item A KMS constraint imposing fluctuation-dissipation relations.
\end{itemize}
When calculating SFFs, one typically sets the external sources $A$ to zero, so the action can be written purely in terms of the derivatives of the $\phi$s.

The $\phi$s have a physical interpretation as phases transforming under the $U(1)$ symmetry. Performing a symmetry operation corresponds to adding a constant to $\phi$. If the symmetry is compact, this requires that adding $2\pi R$ to $\phi$ is actually a gauge transformation that doesn't change the state at all.

As discussed in~\cite{winer2020hydrodynamic}, in order to calculate a spectral form factor, one performs the hydrodynamic integral with periodic boundary conditions in time. For instance, in a system with only energy conservation, the simplest hydro Lagrangian is
\begin{equation}
    L= C \partial_t \phi_a\partial_t \phi_r.
\end{equation}
It is easy to show by taking derivatives with respect to the $A$s that the energy is $E=\int d^d x C \partial_t \phi_r$. So the SFF becomes
\begin{equation}
    \text{SFF}(T,f)=\int \frac {\mathcal D E \mathcal D \phi_a}{2\pi} f^2(E)\exp(-i\int dt \phi_a \partial_t E)
\end{equation}
The modes with nonzero frequencies enforce the conservation law. This leaves us with just an integral over the zero mode.
\begin{equation}
    \text{SFF}(T,f)=\int \frac {d E d \phi_a}{2\pi} f^2(E)  =\frac T{2\pi}\int dE f^2(E),
\end{equation}
the well-known result for GUE systems (for GOE there is an extra factor of two because one can reverse time on one contour relative to the other).

\subsection{Hydro and the Symmetry-Broken Spectral Form Factor}

Before we get into details, let's ask the most basic question: why should we expect spontaneous symmetry breaking to have any effect whatsoever on the spectral hydrodynamic path integral? For systems with spatial extent, SSB allows novel terms in the hydro action\cite{Glorioso_2017}. But it is impossible to write such terms in 0+1d. However, we know from previous sections that in the case of discrete symmetry, SSB has a clear signature in the connected SFF at short times.

The answer is the in the details, and in particular, the boundary conditions. In conventional fluctuating hydrodynamics, when we break our field into $\phi_a$ and $\phi_r$, it is standard practice to consider an overall constant addition to $\phi_r$ as a gauge symmetry. This is because phases aren't observable, only differences in phases like $\phi_a$ or $\phi_r(t_1)-\phi_r(t_2)$ are. But once a symmetry is spontaneously broken, phases do become observable. Water is invariant under a total rotation, but for ice you get a distinct quantum state.

What does this mean for boundary conditions? When calculating the SFF using conventional fluctuating hydrodynamics, it is sufficient to ensure that $\phi_a$ and $\partial_t \phi_r$ are periodic in time. But for SSB SFF, $\phi_r$ itself needs to be periodic. We can think of restoring the symmetry as gauging away $\phi_r$.

Of course, it is itself a phase defined on a circular manifold, and the path integral can wrap around that manifold any integer number of times. This introduces a summation into our calculation. We will handle the details below.

\subsection{The SSB Hydro Action}

We can write the Lagrangian as  
\begin{equation}
L=M\partial_t \phi_a (\partial_t \phi_r+b\partial_t^2\phi_r)+iM\frac{b}{2\beta} (\partial_t \phi_a)^2
\end{equation}
plus terms with more derivatives and/or fields. We show in appendix \ref{app:HydroDerive} that this Lagrangian can be derived as a continuous limit of the discrete $Z_n$ model in the previous section. The nature of spontaneous symmetry breaking means that in order to study it in 0+1d, we need some sort of large-$N$ limit. Thus, in our case, we are justified in dropping nonlinearities and interactions.

\subsection{Hamiltonian Approach}
\label{subsec:hamApproach}

We have formulated the SFF as a path integral involving a hydro-like effective action. In this subsection, we evaluate this path integral by converting to a corresponding Hamiltonian description. In the next two subsections, we give a direct Lagrangian calculation. Again, the contribution of charge to the SFF is 
\begin{equation}
\begin{split}
Z=\int \mathcal D\phi_1\mathcal D \phi_2 \exp(i\int dt L)f(q_1)f(q_2),\\
L=M \partial_t \phi_a (\partial_t \phi_r+b\partial_t^2\phi_r)+iM\frac{b}{2\beta} (\partial_t \phi_a)^2,
\end{split}
\label{eq:hydroPathIntegral}
\end{equation}
where $f$ is some function (perhaps a Gaussian) to regularize over the infinite sum over charges/momenta.

We make use of the fact that a path integral in one dimension is equivalent to a quantum mechanics problem. The Hilbert space is just a suitable space of functions on the configuration space of the path integral, in this case $U(1)^2$.

Viewed as an effective field theory, this path integral arrives from integrating out a number of microscopic modes, but it is also equivalent to a Schrodinger-like evolution on the square of the Goldstone manifold (the imaginary $\phi_a^2$ part of the action corresponds to a non-unitarity for the fictitious Schrodinger dynamics).

To go from a Lagrangian quadratic in velocities to a Hamiltonian quadratic in momenta essentially involves inverting the Lagrangian. It is convenient to convert back to $\phi_1$ and $\phi_2$. Dropping the dissipative term in the equations of motion, the Lagrangian is
\begin{equation}
    L = \frac{M}{2} (\partial_t \phi_1)^2 - \frac{M}{2} (\partial_t \phi_2)^2 + i \frac{M b }{2\beta} (\partial_t \phi_1 - \partial_t \phi_2)^2.
\end{equation}
The corresponding canonical momenta are
\begin{equation}
    \pi_1 = \frac{\partial L}{\partial (\partial_t \phi_1)} = M \partial_t \phi_1 + i \frac{Mb}{\beta} (\partial_t \phi_1 - \partial_t \phi_2)
\end{equation}
and
\begin{equation}
     \pi_2 = \frac{\partial L}{\partial (\partial_t \phi_1)} = - M \partial_t \phi_2 - i \frac{Mb}{\beta} (\partial_t \phi_1 - \partial_t \phi_2).
\end{equation}
The Hamiltonian is then
\begin{equation}
    H = \pi_i \partial_t \phi_i - L = \frac{\pi_i^2}{2M} - \frac{\pi_2^2}{2M} - i \frac{b}{2\beta M} (\pi_1 + \pi_2)^2.
\end{equation}

Since each $\phi_i$ is periodic with period $2\pi R$, the eigenvalues of $\pi_i$ are quantized in terms of integer charges $q_i$ to be $\frac{q_i}{R}$. Hence, the eigenstates of the Hamiltonian are
\begin{equation}
\begin{split}
H\psi_{q_1,q_2}=E_{q_1,q_2}\psi_{q_1,q_2},\\
E_{q_1,q_2}=(q_1^2-q_2^2)\frac{1}{2MR^2}-i\frac{b}{2\beta M R^2}(q_1+q_2)^2,\\
\psi_{q_1,q_2}(\phi_1,\phi_2)=e^{iq_1\phi_1/R}e^{iq_2\phi_2/R}.
\end{split}
\label{eq:Eexpression}
\end{equation}
The non-decaying solutions correspond to $q_1=-q_2, E_{q_1,q_2}=0$.  At long times, these are the only remaining contributions, and the long-time SFF is just equal to the number of charge sectors allowed in the sum. At general times, the overall sum is
\begin{equation}
    Z=\tr \left[f(q_1)f(q_2)e^{-iHT}\right]=\sum_{q_1,q_2}f(q_1)f(q_2)\exp(-iE_{q_1,q_2}T).
    \label{eq:abCoeff}
\end{equation}

It is worth taking a moment to point out that in the large-$N$ limit necessary for 0d spontaneous symmetry breaking, the fictitious $E$s in expressions \eqref{eq:Eexpression} and \eqref{eq:abCoeff} are $1/N$ quantities, meaning the slowest decay time is actually extensive in the system size.

\subsection{Lagrangian Approach}

In this section, we will start be developing technology for arbitrary Gaussian SFFs. We have our path integral of $\exp\left(i \int \phi_a D\phi_r+\phi_a D_{aa}\phi_a \right)$, where $D=a \prod_{k=1}^{K} (\partial_t-\lambda_k)$, with $\textrm{Re}(\lambda_k)<0$ for all $k$, is some differential operator. The form of the action means that the determinant depends only on $D$, and we will set $D_{aa}$ to zero throughout this section.  

We proceed via spacetime discretization, and say that $\mathcal D\phi_a\mathcal D \phi_r=\prod_{j=0}^{T/\Delta t-1} \frac c{2\pi} d\phi_{aj}d\phi_{rj}$, the factor $c$ in the measure is to be determined. We compute the path integral by going to the Fourier basis
\begin{equation}
    \tilde{\phi}_i(\omega) = \sum_{j} \frac{e^{i \omega t_j}}{\sqrt{T/\Delta t}} \phi_i(t_j),
\end{equation}
where $t_j = j \Delta t$ and $\omega = 2\pi n/T$ for $n=0,\cdots,T/\Delta t -1$. The action becomes
\begin{equation}
    \sum_j \Delta t \phi_a(t_j) [D \phi_r](t_j) = \sum_\omega \Delta t \tilde{\phi}_a(-\omega) a \prod_{k=1}^K \left( \frac{e^{- i \omega \Delta t} -1}{\Delta t} - \lambda_k\right) \tilde{\phi}_r(\omega).
\end{equation}
With only the $ra$ cross terms, the $\tilde{\phi}(-\omega)$ integral just gives $2\pi$ times a delta function of $\Delta t  D \tilde{\phi}_r(\omega)$. 

Hence, the path integral evaluates to 
\begin{equation}
   Z= \prod_{\omega} \frac{c}{a \Delta t} \prod_{k=1}^{K} \left(\frac{e^{-i \omega \Delta t}-1}{\Delta t}-\lambda_k \right)^{-1}.
\end{equation}
What constant $c$ in the measure keeps the integral from blowing up? This is pretty clearly a UV question. We want some $c$ such that
\begin{equation}
Z =\prod_{n=0}^{T/\Delta t-1}  \frac{c}{a \Delta t} \prod_{k=1}^{K} \left(\frac{e^{-2\pi i n \Delta t/T}-1}{\Delta t}-\lambda_k \right)^{-1} = \prod_{n=0}^{T/\Delta t-1}  \frac{c}{a \Delta t^{1-K}} \prod_{k=1}^{K} \left(e^{-2\pi i n \Delta t/T} - 1 -\lambda_k \Delta t \right)^{-1}
\end{equation}
doesn't blow as $\Delta t \rightarrow 0$. If we switch the order of the two products we get 
\begin{equation}
Z=\left( \frac{c}{a \Delta t^{1-K}}\right)^{T/\Delta t} \prod_{k=1}^{K} \left(1-e^{\lambda_kT}\right)^{-1}
\end{equation}
For this not to blow up, we need to regularize with $c=a \Delta t^{1-K}$.

Now that we have a general technology for regularizing CTP integrals, let's apply it to the specific case of the hydro Lagrangian $L = M\partial_t \phi_a (\partial_t \phi_r+b\partial_t^2\phi_r)+iM\frac{b}{2\beta} (\partial_t \phi_a)^2$. The corresponding differential operator is
\begin{equation}
D= - M b \partial_t^2 (\partial_t + b^{-1}).    
\end{equation}
In this case, $a= - Mb$ (although only the magnitude of $a$ is relevant), $\lambda_1 = 0$, $\lambda_2 = 0$, and $\lambda_3 = -b^{-1}$. However, there are two things keeping the hydro path integral from being a vanilla Gaussian integral: the zero-frequency modes where one integrates over the hydro  manifold, and the topological aspect of trajectories that wind around the manifold.  For now, we will focus on the saddle point with no winding.

In the non-winding sector, the non-zero-frequency modes are treated as Gaussian variables, while the integral zero frequency modes are integrated over exactly. The zero modes $\tilde{\phi}_i(0)$ range over a period of length $2\pi R\sqrt{T/\Delta t}$, and there are two of them. The path integral in the order parameter sector is thus
\begin{equation}
\begin{split}
Z_{\text{no-winding}}=\frac {c}{2\pi}(2\pi R)^2 \frac{T}{\Delta t} \prod_{n=1}^{T/\Delta t-1} \frac{c}{a \Delta t^{1-K}} \prod_k \left(e^{-2\pi i n\Delta t/T}-1-\lambda_k \Delta t \right)^{-1}.
\end{split}
\end{equation}
The $\lambda_1=\lambda_2 =0$ terms in the product over $k$ give $\left( \frac{\Delta t}{T} \right)^2$, and the $\lambda_3=-1/b$ term gives $\frac{\Delta t}{b(1-e^{-T/b})}$. These pieces combine to give
\begin{equation}
    Z_{\text{no-winding}}=\frac{c}{2\pi} \frac{\Delta t}{T} (2\pi R)^2 \frac {\Delta t}{b(1-e^{-T/b})}.
\end{equation}
After inserting in our expression for $c=Mb\Delta t^{-2}$ we have 
\begin{equation}
Z_{\text{no-winding}}=\frac {1}{2\pi}\frac{M}{T}(2\pi R)^2 \frac{1}{(1-e^{-T/b})}.
\label{eq:coeffProd}
\end{equation}
If we wanted, we could add in higher derivatives to our original action. This would just result in more decaying terms corrections like the $1-e^{-T/b}$ in the denominator.

But there is a complication to equation~\eqref{eq:coeffProd}! The periodicity of $\phi$ actually means that there is an infinitude of saddle-points. In particular, the saddle-points of the action are parameterized by $n_1,n_2$,  winding numbers of $\phi_1,\phi_2$ around the circle. As a function of $n_r=\frac {n_1+n_2}2,n_a=n_1-n_2$, the winding contribution to the action is
\begin{equation}
\Delta S_{\text{winding}} (n_r,n_a)=M \frac{(2\pi R)^2}{T}\left( n_an_r+\frac{ib}{2\beta}n_a^2\right)
\end{equation}
The full path integral is obtained by summing over these modes, and since each winding sector has the same Gaussian part, we get
\begin{equation}
Z=\frac {1}{2\pi}\frac{M(2\pi R)^2}{T(1-e^{-T/b})}\sum_{n_r,n_a}\exp\left(i M\frac{(2\pi R)^2}{T}n_an_r-M\frac{(2\pi R)^2}{T}\frac{b}{2\beta}n_a^2\right).
\label{eq:WeirdSum}
\end{equation}

\subsection{Dealing With the Discrete Sum}

Equation~\eqref{eq:WeirdSum} is a divergent sum that needs to be regulated. In particular, there are an infinite number of $n_a=0$ solutions that add up. To regularize them, we need to remember that $n_r$ is proportional to the momentum/charge of the system, and so to get a finite enhancement we should only be summing over a finite set of charges, rather than all charges from $-\infty$ to $\infty$ (or the allowed set of microscopic charges). Then we can evaluate the path integral with an $f(Q)$ insertion, where $f$ is a regulating function. (For now we will assume it can be naively inserted and removed at will, and will analyze it more carefully in the next subsection.)
We get
\begin{equation}
Z=\frac {1}{2\pi}\frac{M(2\pi R)^2}{T(1-e^{-T/b})}\sum_{n_r,n_a}f(Q_1)f(Q_2)\exp\left(iM\frac{(2\pi R)^2}{T}n_an_r-M\frac{(2\pi R)^2}{T}\frac{b}{2\beta}n_a^2\right).
\end{equation}
Here charge $Q_i$ is related to $n_i$ by
\begin{equation}
Q_i=2\pi M  R^2 n_i/T.
\end{equation}
Moreover, in writing Eq.~\eqref{eq:WeirdSum} we implicitly assumed that $f$ is slowly varying so that the Gaussian part of the path integral is not appreciably modified.

Note that this expression for $Q$ in terms of winding numbers is such that, if we use it inside the path integral with an $f(Q)$ insertion, then we recover the correct sum over discrete charges of $f(Q)$ in the final answer. However, the quantization of $Q$ in the path integral arising from the winding sectors is different than the microscopic quantization. Either way of doing the sum, direct microscopic calculation or path integral calculation, will give the same answer, and they are related, roughly speaking, by a Poisson resummation. The situation is similar to the comparison between the Hamiltonian and Lagrangian descriptions of a particle on a ring.

For long times, we can rewrite $n_a$ in terms of $Q_a$ to see that the $\propto - Q_a^2 T$ term in the exponent suppresses $Q_a \neq 0$ contributions. The path integral is thus
\begin{equation}
\frac {1}{2\pi}\frac{M(2\pi R)^2}{T(1-e^{-T/b})}\sum_{n_r}f^2(2\pi M  R^2 n_r/T)=\int dQ f^2(Q),
\end{equation}
exactly what we want. Again, this integral over charge $Q$ approximates the discrete sum on the left hand side and is not a sum over the microscopic charges, but in the limit of slowly varying $f$ where Eq.~\eqref{eq:WeirdSum} is valid, we obtain the same answer.

Note that the functional substitution $Q=M\partial_t \phi_r$ would give us the exact same action as in the non-SSB case. The reason we get a more complicated short-time behavior is that we are evaluating the path integral in a different way. One way to look at the change is as a change in boundary conditions. Before breaking the symmetry, we treated an overall addition to $\phi_r$ as a gauge symmetry, and allowed arbitrary additions of it over the period. Whereas now we require that $\phi_r$ change by a quantized amount equal to $2\pi R n_r$.

\subsection{A More Careful Accounting of $f$}

We can express the filter function as 
\begin{equation}
 f(q) = \int_{0}^{2\pi R} dx \tilde f(x)e^{iqx}.   
\end{equation}
Inserting any operator $f(Q)$ into the path integral is thus equivalent to a double integral over basic insertions of the form $e^{i Q_1 x_1} \otimes e^{ i Q_2 x_2}$. Since $Q_i$ is conjugate to $\phi_i$, this insertion when acting on a state of definite $\phi_i$ shifts the value by $ - x_i$. When placed after the final resolution of the identity in the path integral, one gets a delta function $\delta(\phi_i(0) - \phi_i(T)+x_i)$ that effectively shifts the boundary condition to $\phi_i(T)=\phi_i(0)+x_i+2\pi R n_i $. The path integral \eqref{eq:hydroPathIntegral} can thus be written as
\begin{equation}
\begin{split}
Z=\int dx_1dx_2 \int \mathcal D_{x_1}\phi_1\mathcal D_{x_2} \phi_2 \exp(i\int dt L)\tilde f(x_1)\tilde f(x_2),\\
L=M \partial_t \phi_a (\partial_t \phi_r+b\partial_t^2\phi_r)+iM\frac{b}{2\beta} (\partial_t \phi_a)^2,
\end{split}
\label{eq:hydroPathIntegralBC}
\end{equation}
where $\mathcal D_{x}$ is the path integral measure with twisted periodic boundary conditions $\phi_i(T)=\phi_i(0)+x_i+2\pi R n_i$. 

If we repeat the analysis in the last two sections, we find that these twisted boundary conditions don't change the Gaussian part of the path integral at all. Defining $x_r=\frac{x_1+x_2}{2}$, $x_a=x_1-x_2$, equation \eqref{eq:WeirdSum} becomes 

\begin{equation}
\begin{split}
Z=&\frac {1}{2\pi}\frac{M(2\pi R)^2}{T(1-e^{-T/b})}\int_{0}^{2\pi R} dx_1\int_{0}^{2\pi R}dx_2\sum_{n_r,n_a}\tilde f(x_1)\tilde f(x_2)  \exp( \Delta S) \\
& \text{with } \Delta S = \left(iM\frac{\left((2\pi R)n_r+x_r\right)\left((2\pi R)n_a+x_a\right)}{T}-M\frac{\left((2\pi R)n_a+x_a\right)^2}{T}\frac{b}{2\beta}\right),\\
Z =&\frac {1}{2\pi}\frac{M(2\pi R)^2}{T(1-e^{-T/b})}\int_{-\infty}^{\infty} dx_1\int_{-\infty}^{\infty} dx_2\tilde f(x_1)\tilde f(x_2)\exp\left(iM\frac{x_rx_a}{T}-M\frac{x_a^2}{T}\frac{b}{2\beta}\right),
\end{split}
\label{eq:WeirderSum}
\end{equation}
where in the last equation we extend $\tilde f$ to be a period function with period $2\pi R$. 
This formula is fully general for any regulating function $f$ no matter how rapidly it varies. 

Let's investigate its behavior at short times. At short times we can consider the $\tilde f$s roughly constant compared to the Gaussian integral. Performing this integral and factoring in the determinant $2\pi \frac{T}{M}$ from the delta function, we have the short-time formula
\begin{equation}
    Z_{\textrm{short time}}=\frac{1}{(1-e^{-T/b})}(2\pi R)^2 \tilde f(0)^2=\left(\sum_{-\infty}^\infty f(q)\right)^2.
\end{equation}

More generally, we can replace $\tilde f(x_1)=\frac 1{2\pi R}\sum_{q'_1}e^{-iq'_1x_1}$. If we do the same for $x_2$, and define $q'_r=\frac{q'_1+q'_2}{2},$ $q'_a=q'_1-q'_2$, we have a Gaussian integral that works out to
\begin{equation}
    Z=\frac{1}{(1-e^{-T/b})}\sum_{q'_1,q'_2}f(q'_1)f(q'_2) \exp(i\frac{T}{M}q'_aq'_r+\frac{2b}{\beta}\frac{T{q'}_a^2}{M}).
\end{equation}
If we make the $T\gg b$ assumption of subsection \ref{subsec:hamApproach}, this is the exact same formula as \eqref{eq:abCoeff}. For long $T$ the $q'_a\neq 0$ terms become zero and we have 
\begin{equation}
    Z_{\textrm{long time}}=\sum_{q'}f^2(q').
\end{equation}

\section{Non-Abelian Hydro and the SFF}
\label{sec:NonAbelianHydro}

In this section, we treat the case of spontaneous symmetry breaking of continuous non-Abelian symmetries. As in the continuous Abelian case, a modified version of the hydro theory of non-Abelian SSB provides a useful way to formulate the SFF. 

\subsection{Overview of Non-Abelian Hydro}

In Abelian $U(1)$ hydrodynamics, the phase field $\phi$ can be regarded as an element of the group $U(1)$ by exponentiating it. In non-Abelian hydrodynamics~\cite{nonAbelian}, we replace the exponentiated phase field with a field $g$ that takes values in the (non-Abelian) global symmetry group $G$. The hydro generating function is expressed as 
\begin{equation}
\begin{split}
    Z_{\text{hydro}}[A^\mu_1(t,x),A^\mu_2(t,x)]=\int \mathcal D g_1\mathcal D g_2 \exp\left(i\int dt d^d x W[B_{1\mu},B_{2\mu}]\right),\\
    B_\mu=A_\mu + i g^{-1}\partial_\mu g,
\end{split}
\end{equation}
a non-Abelian generalization of Eq.~\eqref{eq:AbelianB}. This action and generating function will always have a $G$ premultiplication symmetry. Note that in the case of $G=U(1)$, we recover the original Abelian hydro formalism.
For the purposes of this paper we will focus on the unsourced $A=0$ case, and also restrict to zero spatial dimensions. 

Note that any action constructed out of $B$s will have a $G$ premultiplication symmetry for both the right and left replica, for a total symmetry group of $G_L\times G_R$. In conventional hydro, this symmetry is broken to the diagonal by the future and past boundary conditions, but these boundary conditions are not present in the SFF case. A representative Lagrangian for $0$d non-Abelian hydro would be
\begin{equation}
L=B_{ta}(M+Mb\partial_t)B_{tr}+iM\frac{b}{2\beta}B_{ta}^2
\label{eq:nonAbelianAc}
\end{equation} 
where the $B$s are in the adjoint representation of the Lie algebra of $G$, and there is an implied summation over the representation indices.

Action \ref{eq:nonAbelianAc} has two time derivatives. Since we are working in 0+1d QFT, this means there is a quantum mechanics interpretation, just as in the Abelian case there was an interpretation in terms of non-unitary evolution on $U(1)^2$.

\subsection{Hamiltonian Approach for Full SSB}

We can use a Legendre transform to go from the Lagrangian \eqref{eq:nonAbelianAc} to a Hamiltonian,
\begin{equation}
H_{\text{doubled}}=\frac{1}{2M}\sum L_{i1}^2-\frac{1}{2M}\sum L_{i2}^2-i\frac{b}{2\beta M}\left(\sum L_{i1}+L_{i2}\right)^2,
\end{equation}
where we sum over group generators $L_i$, which are canonically conjugate to the velocity components of $i g^{-1} \partial_t g$. The sum $\sum_i L_i^2$ is called the Casimir operator $L^2$, and has a number of important properties. Intuitively, it plays the same role as a Laplacian, but on a group manifold. Just as the Laplacian operator commutes with any momentum operator, $L^2$ commutes with all elements of the group $G$. As such it can be shown to be constant within any irreducible representation of $G$. For the Abelian case $G=U(1)$, where irreducible representations $R$ are parameterized by integer charges $q$, $L^2(R)=q^2$.

Let's evaluate
\begin{equation}
    Z=\tr \left[ e^{-iH_{\text{doubled}}T}f(L^2_1)f(L^2_2) \right],
\end{equation}
where we promoted the filter function $f(Q)$ from the Abelian case to $f(L^2)$ in the non-Abelian case. Wavefunctions transform in the square of the regular representation of $G$. This representation means that each replica has $|R_i|$ copies of representation $R_i$. The Hamiltonian depends on the Casimir of each $R_i$ and on the composite system Casimir $(L_1+L_2)^2$.  The enhancement factor is thus
\begin{equation}
Z=\sum_{R_1,R_2,\bar{R}} f(L^2(R_1))f(L^2(R_2))|R_1||R_2||\bar R|\exp\left(- i \frac{[L^2(R_1)-L^2(R_1)]T}{2M} -\frac{b L^2(\bar R)T}{2\beta M}\right)n(\bar R)_{R_1R_2},
\label{eq:tripleSum}
\end{equation}
where $n(\bar R)_{R_1R_2}$ is the number of times $\bar R$ appears in $R_1\times R_2$.  For instance if $R_1$ and $R_2$ are the spin 1/2 and spin 1 representations of $SU(2)$ and $\bar R$ is the spin 1/2 representation, then $n(\bar R)_{R_1 R_2}=1$.

The longtime behavior is given by the $\bar R$ trivial case where $L^2(\bar R)=0$. This requires $\bar R$ to be the trivial 1D representation. If $\bar R$ is trivial,  $n(\bar R)_{R_1R_2}$ is zero unless $R_1$ is the complex conjugate of $R_2$,  in which case it is one. So the long time value is 
\begin{equation}
Z(T\rightarrow \infty)=\sum_{R_1} f^2(L^2(R_1))|R_1|^2 .
\end{equation}
This is exactly what one would predict from random matrix theory. We can also make use of $|R_1||R_2|=\sum_{\bar R}n(\bar R)_{R_1R_2} |\bar{R}|$ to show that in the short time limit this is
\begin{equation}
   Z(T \rightarrow 0) =\sum_{R_1}\sum_{R_2} f(L^2(R_1))f(L^2(R_2))|R_1|^2|R_2|^2.
\end{equation}
Since the regular representation has $|R_i|$ copies of $|R_i|$, this is essentially saying all states of all charges constructively interfere.

\subsection{Partial SBB}

If a subgroup $G'$ of $G$ is unbroken, this means overall $G'$ transformations that affect both replicas are unphysical. Thus we should gauge them out. In general, one gauges out a group by inserting a projector
\begin{equation}
P=\frac 1 {\textrm{Vol }G'}\int_{G'} dg' g'
\label{eq:projection}
\end{equation}
into the path integral. If we break representations $\bar R$ of $G$ into representations of $G'$, this projects out all nontrivial representations of $G'$. For instance if $G'=G$, this will remove all the nontrivial $R'$ leaving us with the result for RMT with symmetry group $G$. If $G'$ is trivial, nothing is projected out and equation \eqref{eq:tripleSum} still holds. In general, different symmetry-breaking patterns can be thought of as different degrees of freedom being observable, which means different projection operators $P$ are needed as boundary conditions.

For example, let's imagine a system with $SO(3)$ symmetry, which is broken down to $SO(2)$ by an order parameter. We are interested in some term in equation \eqref{eq:tripleSum}, for example $R_1=\textbf 5, R_2=\textbf 3, \bar R=\textbf 3$. We can verify that $n(\bar R)_{R_1R_2}=1$, meaning that when we multiply our representations we get one copy of $\bar R=\textbf 3$. What happens to this representation under the projection in equation \eqref{eq:projection}? To answer this, let's break of the representation of $G$ into representations of $G'$. The vector representation of $SO(3)$ breaks into a scalar and vector representation of $SO(2)$. Integrating $g'$ over the vector representation (as over any nontrivial representation of any group) we get zero. Integrating over the trivial representation we of course get one. So the projection operator projects down from three dimensions to one, and the $|\bar R|$ in equation \eqref{eq:tripleSum} becomes 1 instead of 3.

\section{Discussion}
In this paper we extended the understanding of quantum chaotic level repulsion to include systems with spontaneous symmetry breaking. We started with toy models with discrete symmetries, solved them, and confirmed our solutions with exact diagonalization. Next we used hydrodynamics, extending known techniques for unbroken symmetry to the case of spontaneous breaking. The technique is powerful enough to prove the correct longtime behavior, and flexible enough to handle any possible symmetry breaking pattern. Interestingly, we found that SSB typically enhances the SFF beyond that of a system with unbroken symmetry. In terms of the spectral form factor we have schematically
\begin{equation}
    \textrm{No Symmetry}<\textrm{Symmetry}<\textrm{Spontaneously Broken Symmetry}.
\end{equation}
One is left wondering how gauge symmetry might fit into that hierarchy.

In terms of SSB, the next step would be to handle higher dimensional systems. At least two interesting phenomena would reveal themselves in this case. First, the presence of sound poles associated with Goldstone modes. In higher dimension spontaneous symmetry breaking allows new terms in the hydrodynamic Lagrangian consistent with unitarity, such as $\phi_a\partial_x^2\phi_r$. This would allow the hydrodynamic variables to have sound poles, leading to a potentially rich new phenomenology in the SFF.

Higher dimensions also allows the possibility of topological effects. For instance, in a periodic system the Goldstone mode could wrap around the manifold several times. This new topological charge would lead to an expansion in the number of sectors and an additional enhancement to the SFF that could last for exponentially long times until the system tunnels into a topologically uncharged state. More exotic Goldstone manifolds and spatial manifolds would result in even more exciting topological concerns.

Finally, there is the issue of the plateau structure, entirely ignored in this paper. The lack of the hydrodynamic description of plateau behavior is made all the more striking by the fact that certain systems with some sort of resonant behavior (the peaks in figure \ref{fig:ZnMomentum}) can have `ramp' values of the SFF exceeding the final plateau value.

\section{Acknowledgements}

This work is supported in part by the Simons Foundation via the It From Qubit Collaboration (B. S.) and by the Air Force Office of Scientific Research under award number FA9550-17-1-0180 (M.W.). M.W. is also supported by the Joint Quantum Institute.

\appendix

\section{Appendix: Internal Charge}
\label{app:IntCharge}

For many realistic systems, there is charge/momentum contained within the state $\psi$,  not just encoded in the pattern of $\phi$s.  A nice example of this is a superconductor. There is some charge in the condensate,  but plenty of other charge in the system,  including in uncondensed electrons and atomic nuclei. Likewise,  if we look at a mechanical example of rotational SSB like a planet with a cloud of moons orbiting a star, there is some orbital angular momentum of the overall system,  but the system of planets and moons has its own intrinsic angular momentum.  What sort of Hamiltonian captures this situation?

Let's say the states are characterized by an order parameter $\phi$, an internal angular momentum $L$, and an internal state $\ket \psi$. 
The simplest $G$-invariant Hamiltonian we can write for this is given by equation \eqref{eq:HRandom} for with internal-$G$-invariant $H_0$ and $H_i$s. This candidate $H$ actually has two copies of the symmetry, an internal one and an external one. 

To make the model more realistic, we need to have some coupling which breaks us down to just one copy of the symmetry. First let's do it for $Z_n$. We can parameterize the sectors by order parameter $0\leq \phi <n$ and $q$. Then we have a coupling connecting the $\phi,q$ and $\phi,q'$ sectors proportional to $e^{2\pi i /n \phi(q-q')}$.  This obviously doesn't commute with bland translation or bland conservation of $q$.  But it does retain an overall $Z_n$ symmetry.
Written in block matrix form, such a Hamiltonian for a $Z_4$ symmetry might take the form
\begin{equation}
H=\begin{pmatrix}
H_0^0&I&0&I\\
I&H_0^1&I&0\\
0&I&H_0^2&I\\
I&0&I&H_0^3
\end{pmatrix}
\end{equation}
Where $H_0^\phi$ is, itself, a block matrix of the form
\begin{equation}
H_0^\phi=\begin{pmatrix}
Q_{00}&e^{2\pi i/4\phi(0-1)}Q_{01}&e^{2\pi i/4\phi(0-2)}Q_{02}&e^{2\pi i/4\phi(0-3)}Q_{03}\\
e^{2\pi i/4\phi(1-0)}Q_{10}&Q_{11}&e^{2\pi i/4\phi(1-2)}Q_{12}&e^{2\pi i/4\phi(1-3)}Q_{13}\\
e^{2\pi i/4\phi(2-0)}Q_{20}&e^{2\pi i/4\phi(2-1)}Q_{21}&Q_{22}&e^{2\pi i/4\phi(2-3)}Q_{23}\\
e^{2\pi i/4\phi(3-0)}Q_{30}&e^{2\pi i/4\phi(3-1)}Q_{31}&e^{2\pi i/4\phi(3-2)}Q_{32}&Q_{33}
\end{pmatrix}
\end{equation}
While the $Q_{ii}$ matrices all have to be square, Hermitian matrices and $Q_{ij}=Q_{ji}^\dagger$, in general the internal-charge-i subspace and the internal-charge-j subspace can be of totally difference sizes.

If we have a more general group,  the elements are parameterized by order parameter $\phi$,  irrep $R$ and vector $k$ within that irrep.  We connect $\phi,R$ and $\phi,R'$ by $R_{G(\phi)}MR'_{G^{-1}(\phi)}$ where $G(\phi)$ is a group element that gets us to $\phi$, and $M$ is invariant under the unbroken part of the symmetry.

\section{Appendix: Long-Time Behavior with Non-Abelian Discrete Groups}
\label{app:longTime}

The enhancement factor for non-Abelian discrete symmetry groups is $\tr[e^{-\text{Trans}(E)T}]$ with transfer matrix
\begin{equation}
\begin{split}
 \text{Trans} =\frac 12 \sum r_i (M_i\otimes I-I\otimes M_i^T)(M_i^T\otimes I-I\otimes M_i)+(M_i^T\otimes I-I\otimes M_i)(M_i\otimes I-I\otimes M_i^T).
\end{split}
\end{equation}
In this section, we count the zero modes of the transfer matrix. Such a zero mode must be annihilated by $M_i \otimes I-I\otimes M_i^T$ for all $M$s, which is a heavy constraint. 

For our analysis, we will need to decompose a vector space $C^{\Phi}$ into irreps. This space is a representation of $G$, with the $G$ matrices forming permutation matrices which permute the elements of $\Phi$ according to the group action. The matrices $M_i$ are matrices acting on this space which commute with every element of $G$. We can decompose the representation $C^\Phi$ into irreducible representations of $R$ as
\begin{equation}
    C^{\Phi}=\bigoplus_R R^{\oplus K_R},
\end{equation}
where each representation $R$ appears $K_R$ times. A vector $v$ in $C^\Phi$ can be written as $v_{R,k,\mu}$, where $R$ denotes the irreducible representation it transforms in, $0\leq k<K_R$ indicates which copy of $R$, and $\mu$ is the index within the representation $R$. In this case, the requirement that the $M_i$s commute with elements of $G$ mean that they can act only on the $k$ index, in a way not depending on the $\mu$ index. If we sum over a large enough collection of $M$s (in particular, enough $M$s so that their action on the $k$ indices don't all commute) then the only vectors on $C^{\Phi\otimes \Phi}$ which are annihilated by the transfer matrix are ones in which the left and right $k$ indices are maximally entangled. We are free to choose the represenation $R$ and the indices $\mu_1$ and $\mu_2$ for the right and left replicas. So we have
$\sum_R |R|^2$ zero modes. This is also the random matrix theory prediction for the long-time ramp enhancement.

\section{Appendix: Deriving Hydro from the Discrete Case}
\label{app:HydroDerive}

Let's talk about how a term like equation \eqref{eq:transfer2} gives rise to a Goldstone-like theory.  It can be thought of as decoupled diffusion on both the right and left copies of the system,  with positive and negative imaginary diffusivities.  Of course, this is just QFT with a canonical kinetic term
\begin{equation}
L=\frac 12\phi_1\partial_t^2\phi_1-\frac 12\phi_2\partial_t^2\phi_2=\phi_a\partial_t^2\phi_r.
\label{eq:Lra}
\end{equation}
This is already similar to the Goldstone theory, but it needs an $aa$ term. This comes from the contribution of the matrix in equation \eqref{eq:transferEq}. This matrix can be thought of as generating an un-intuitive sort of random walk on $\Phi^2$.  The most un-intuitive part is that the transfer probabilities aren't all positive. We can see from the formula that there is a morass of positive and negative signs. In order to build intuition, let's consider the case of $Z_n$ symmetry acting on $n$ elements. We know that the matrices $M_i$ are indexed by $\Phi^2/G$, which in this case is just the collection of jump sizes,  ranging from the trivial jump to jumping $n-1$ to the right. For simplicity, let's just look at $(M_1)_{ij}=\delta_{i,j+1}$,  the nearest-neighbor jump. 

What sort of transfer matrix does this give rise to? Remembering that the transfer matrix is a linear map from the vector space $R^{\Phi^2}$ to itself and thus has a total of 4 $\Phi$ indices,  it is unilluminating to try to write the whole thing.  But we can write that it is 
\begin{equation}T_{ij,i'j'}\propto 2\delta_{i,i'}\delta_{j,j'}-\delta_{i,i'+1}\delta_{j,j'+1}-\delta_{i,i'-1}\delta_{j,j'-1}
\end{equation}
If we go from a discrete $Z_n$ to a continuous $U(1)$,  this transfer matrix corresponds to $\phi_r$ undergoing diffusion while $\phi_a$ doesn't change at all.  When we combine this with the transfer matrix in equation \eqref{eq:transfer2} we add the covariances/correlators. Equation \eqref{eq:Lra} only has $ar$ correlators. Adding in an $rr$ correlator results in a Lagrangian 
\begin{equation}
L=\phi_a\partial_t^2\phi_r+iC\phi_a\partial_t^2\phi_a.
\end{equation}
This is a generic hydrodynamic action for a superfluid, obtained entirely through taking the continuous limit of the transfer matrices.

\section{Appendix: Time-Reversal Symmetry}
\label{app:TR}

Let's consider spontaneous time-reversal symmetry breaking. First, what is a good RMT-like toy model of the phenomenon? Consider a block Hamiltonian of the form 
\begin{equation}
    H=\begin{pmatrix}H_0&H_1\\H_1&H_0^*\end{pmatrix},
    \label{eq:trmodel}
\end{equation} 
where $H_0$ is an $N\times N$ GUE matrix and whose real and imaginary parts have variance $J^2/N$, and $H_1$ is a randomly selected GOE matrix whose elements are all independent and have variance $J_1^2/N$. This matrix has an anti-unitary time reversal symmetry which conjugates the elements and switches the two blocks (which we will label the $+$ and $-$ blocks). But within each block, there is no time-reversal symmetry. Since the system will choose a block and only slowly tunnel back and forth, we say that the system spontaneously breaks the time-reversal symmetry.

We can study this using a similar instanton-like approach as with $Z_2$ SSB. We start with a doubled system.  Since one copy of the system has two sectors, a pair of copies has $2\times 2=4$ sectors.  If the system is in the $++$ or $--$ sectors, it has exactly the same transfer matrix as a GUE matrix with two sectors.  The processes contributing are the $++\to--/--\to++$ process with amplitude given by Fermi's golden rule, and the negative-amplitude $++\to++/--\to--$ processes. This gives an enhancement of $\tr e^{MT}=1+e^{-r T}$, (where $r$ is the transition rate) over the GUE result. However, we can also time-reverse the left contour with respect to the right.  Now we have a system starting in either the $+-$ or $-+$ states,  and the left replica is performing a time-reversed version of the right-replica's evolution. This gives another $1+e^{-rT}$ contribution. The overall enhancement factor is thus $2(1+e^{-rT})$ times the GUE ramp, which is of course $1+e^{-rT}$ times the GOE ramp. 

One thing which bears discussion and which is not fully understood is what happens when the Thouless time becomes comparable to the Heisenberg time for individual subsystems, in large part because we don't have a path integral-like picture for how the ramp gives way to the plateau. For GOE-like systems, plateau-like behavior sets in gradually even before the Heisenberg time. One guess then to assume that the ramp behavior for time reversal SSB is thus $(1+e^{-rT})\text{SFF}_{\text{GOE}}(T)$ even when $t_{\text{Thouless}}\lesssim t_{\text{Heis}}$.
\begin{figure}
\centering
    \includegraphics[width=.5\textwidth]{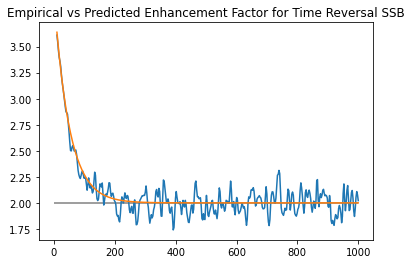}
    \caption{Predicted (orange) and observed (blue) enhancement relative to GOE expectation for the model in equation \eqref{eq:trmodel}.}
    \label{fig:my_label}
\end{figure}
\bibliographystyle{ieeetr}
\bibliography{ssb.bib}
\end{document}